\documentclass[conference,a4paper]{IEEEtran}
\usepackage{xcolor}
\usepackage{balance}
\usepackage{cite}
\usepackage{amsmath,amssymb,amsfonts}
\usepackage{algorithm,algorithmic}
\ifCLASSINFOpdf
  \usepackage[pdftex]{graphicx}
 \else
   \usepackage[dvips]{graphicx}
 
\fi

\ifCLASSOPTIONcompsoc
 \usepackage[caption=false,font=normalsize,labelfont=sf,textfont=sf]{subfig}
\else
 \usepackage[caption=false,font=footnotesize]{subfig}
\fi
\begin{document}
\title{\ Optimization of Super-Directive Linear Arrays with Differential Evolution for High Realized Gain}

\author{\IEEEauthorblockN{
Ihsan Kanbaz\IEEEauthorrefmark{1}\IEEEauthorrefmark{2},   
Okan Yurduseven\IEEEauthorrefmark{1},   
and Michail Matthaiou\IEEEauthorrefmark{1}   
}                                     
\IEEEauthorblockA{\IEEEauthorrefmark{1}
Centre for Wireless Innovation (CWI), Queen’s University Belfast, Belfast BT3 9DT, U.K }
\IEEEauthorblockA{\IEEEauthorrefmark{2}
Department of Electrical and Electronics Engineering, Gazi University, Ankara, Turkey 
\\e-mail: \{i.kanbaz, okan.yurduseven, m.matthaiou\}@qub.ac.uk}
}

\maketitle
\begin{abstract}
Due to the low impedance and high feeding currents, it is naturally challenging to design super-directive antenna arrays that perfectly match the feed line, and this becomes almost impossible as the number of elements increases. In this paper, we assert that it is crucial to consider the trade-off between directivity and overall efficiency (to achieve high realized gain) before employing super-directive arrays in real-world applications. Given this trade-off (high directivity and low mismatch for high realized gain), a 4-element dipole array (unit array) is optimized using the differential evolution (DE) algorithm. Then, the performance of the unit array in subarray configuration scenarios is analyzed. Finally, the obtained parameters are verified using the CST full-wave simulation software. The results clearly indicate that the proposed unit array is a strong candidate for dense array applications, particularly in the context of massive multiple-input multiple-output (MIMO), thanks to its notable high gain and efficiency.

\let\thefootnote\relax\footnotetext{This work was supported by a research grant from the European Research Council (ERC) under the European Union’s Horizon 2020 research and
innovation programme (grant No. 101001331).}

\end{abstract}

\vskip0.5\baselineskip
\section{Introduction}
Although mutual coupling is often seen as a disadvantage, extensive work has been done on leveraging it in order to reduce the physical size of array antennas and increase their directional properties (Directivity, $D$) \cite{harrington,ESA2}. While the maximum directivity is directly proportional to the number of elements in a classical array consisting of isotropic sources ($D \propto N$), it was theoretically shown in the seminal work of Uzkov \cite{uzkov} that if the distance between the elements converges to zero, the directivity of the array can reach up to the square of antenna elements number ($D \propto N^2$) towards the end-fire direction. However, for a considerable number of years, these systems, known as super-directive arrays (which can be broadly defined as antennas whose end-fire directivity exceeds that of uniformly excited apertures \cite{NewDef}), have been deemed impractical due to their susceptibility to sensitive feeding currents and limited efficiency resulting from feed line impedance mismatch \cite{Unidirectional},\cite{OnEndFire}. Thus, there is a significant need to develop optimization frameworks to realize these antenna arrays and provide practical excitation topologies to make super-directive arrays a feasible technology.

Since mutual coupling is minimized in classical arrays (whenever the inter-element distance is chosen more than half a wavelength or reduced by complex techniques), designing a single antenna impedance matching circuit for the array elements is sufficient. However, placing the antennas closer to each other causes impedance mismatch due to factors that are not easy to calculate, such as surface currents and near-field radiations \cite{Towards}. In the scenario where each antenna is simultaneously excited, the array elements will be subject to different mutual coupling, so their input impedances will also differ. Therefore, different matching circuits or techniques must be used for each antenna element compatible with the feed line. In many studies carried out in the literature, it has been suggested that the input impedances of some antenna elements are very low (even negative for some configurations), and, hence, impedance matching becomes almost impossible \cite{SuperDirectiveAntennas}. Consequently, this inherent incompatibility is of paramount significance, representing a formidable challenge in super-directive arrays alongside the existing constraints.

We now recall that there is a close relationship between directivity and gain. The gain expression is obtained if the radiation efficiency is included in the directivity term \cite{balanis}. Another gain expression that needs attention is the realized gain which also includes the line impedance mismatch. In a perfectly matched array, the overall efficiency, encompassing both line and radiation efficiencies, tends to be exceptionally high. Many studies commonly rely on this assumption while neglecting the concept of realized gain \cite{OptCurrent, Maximumdirectivity, MultiOpt, Tomas2020}. However, in super-directive arrays, the realized gain diminishes considerably due to impedance mismatches within the array's transmission lines \cite{ESA2,SuperDirectiveAntennas}. Therefore, enhancing the realized gain requires utilizing a multi-parameter optimization approach involving adjustments to element spacing, antenna feed currents, and antenna dimensions. Although there are limited studies on the realized gain (e.g. \cite{CompEnd,SWE3,ImpedanceMismatch}), to the best of our knowledge, there has yet to be any previous work optimizing the realized gain by an iterative algorithm verified by a full wave electromagnetic simulation program. Against this background, our paper makes the following main contributions:
\begin{itemize}
\item We employ a DE algorithm to showcase the optimum inherent trade-off between high directivity and low impedance mismatch. 
\item We study a more complicated scenario that contains four elements dipole array (unit array) instead of a simple two element configuration. Subsequently, we assess the performance of the unit array in different subarray configurations.
    \item We validate the obtained parameters through a rigorous assessment using comprehensive full-wave electromagnetic simulation software. 
\end{itemize}

The results unequivocally demonstrate that the proposed array, characterized by its high gain and efficiency parameters, represents a valuable and promising choice for dense array applications, particularly for massive MIMO architectures.

\textit{Notation}: The boldface letters stand for vectors and matrices; $\textbf{X}^T$ $\textbf{X}^*$ and $\textbf{X}^H$ represent the transpose, conjugate and conjugate transpose operations, respectively. Finally, $\|{\bf{x}}\|^2 $ denotes the $l_2$ norm of $\bf{x}$.


\section{Theoretical Background of Super-Directive Antenna Arrays}
\label{sec:theory}
A dipole of length $l$ and radius $\rho$, positioned on the $x$-axis, excited in the $z$-direction, is considered for the unit element. The electric field of each dipole in far-field distance ($r$) and observation angle $\theta$ is written as follows:
	\begin{equation}
		{\bf{E}}_n(r,\theta) = -j\eta {\bf{F}}(\theta) \frac{e^{-jkr}}{2\pi r}i_n, \label{eqEfield}
	\end{equation}
	where $\eta\approx120\pi$ is the impedance of free space, $i_n$ is the complex excitation current, $k=2\pi/\lambda$ represents the wave number, $\lambda=\frac{c}{f}$ is the wavelength while $c$, $f$ represent the speed of light and center frequency, respectively. Here, the term $\bf F(\theta)$ is called the element factor and is theoretically independent of the azimuth angle ($\phi$) (for a $z_n$ oriented dipole) and expressed as follows \cite{balanis}:
		\begin{equation}
		\begin{aligned}
			{\bf{F}}(\theta) = \frac{\cos(\frac{kl}{2}\cos\theta)-\cos(\frac{kl}{2})}{\sin(\theta)} \textbf{a}_{{\theta}}.\label{F}
		\end{aligned}
	\end{equation}
Here, $\textbf{a}_{\theta}$ denotes the unit vector in the polar direction. Hence, the unified electric field of the array can be represented as:
 	\begin{equation}
		\begin{aligned}
			\textbf{E}(r,\theta, \phi)=-j\eta \frac{e^{-jkr}}{2\pi r} \sum_{n=0}^{N-1}e^{-jk\bf{r}.\bf{r}_n}i_n\bf{F}(\theta),
		\end{aligned}
	\end{equation}
 where, the position vector $\textbf{r}$ is defined as $(\sin\theta \cos\phi,\sin \theta \sin \phi, \cos \theta)^T$, which points to the observation location. On the other hand, $\textbf{r}_n$ represents the position vector of the $n$-th element and is defined as $(nd,0,0)$. Therefore, the radiation intensity can be calculated to express the total power radiated from the antenna as follows \cite{dovelos}:
		\begin{align}
			U(\theta, \phi) &= \frac{r^2}{2\eta}\|{\bf{E}}(r,\theta, \phi)\|^2 
			\\&= \frac{\eta}{8\pi^2} \|{\bf{F}}(\theta)\|^2 \left|{\bf a}^H (\theta, \phi) \cdot {\bf{i}}\right|^2, \label{eq:U}
		\end{align}
 where ${\bf a}(\theta,\phi) = \begin{bmatrix} e^{-jk\bf{r}.\bf{r}_0}, ... , e^{-jk\bf{r}.\bf{r}_n} \end{bmatrix}^T \in \mathbb{C} ^{N \times 1}$ and ${\bf i} = \begin{bmatrix} i_0, ... , i_n \end{bmatrix}^T \in \mathbb{C} ^{N \times 1}$ are the far-field array response vector and vector input currents, respectively. The total radiated power is obtained by integrating the radiation intensity as follows \cite{Orfanidis}:
 \begin{equation}
		\begin{aligned}
		P_\mathrm{rad} = \frac{1}{2} {\bf{i}}^H \cdot \Re{\{\bf{Z}\}'} \cdot {\bf{i}}, \label{Prad}
		\end{aligned}
	\end{equation}
where $\Re\{\cdot\}$ denotes the real part operation, while ${\bf Z}' \in \mathbb{C} ^{N \times N}$ represents the input impedance matrix of a lossless dipole array. In a practical context, assessing the conductor losses of the dipoles becomes imperative. The computation of the loss resistance ($R_\mathrm{loss}$) for a current-carrying conductor wire can be carried out following the methodology of \cite{dovelos}. Therefore, the overall power loss is calculated as follows:
	\begin{equation}
		P_\mathrm{loss} = \frac{1}{2}R_\mathrm{loss}{||\bf{i}}||^2 =\frac{1}{2} {\bf{i}}^H \cdot {\bf R}_\mathrm{loss} \cdot {\bf i}, \label{eqPloss}
	\end{equation}
    where ${\bf R}_\mathrm{loss} =R_\mathrm{loss} {\bf{I}}_N \in \mathbb{R} ^{N \times N}$, where ${\bf{I}}_N$ represents an $N \times N$ identity matrix. From this perspective, the overall input power at the antenna feed point is computed in the following manner:
    \begin{align}
P_\mathrm{in} &= P_\mathrm{loss} + P_\mathrm{rad}
			&= \frac{1}{2} {\bf{i}}^H \cdot {\bf R}_\mathrm{loss} \cdot {\bf i} + \frac{1}{2} {\bf{i}}^H \cdot \Re{\{\bf{Z}\}'} \cdot {\bf{i}}, 
   \label{eq:Pin}
	\end{align}
	where
	\begin{equation}
		\Re{\{\bf{Z}\}} \overset{\Delta}{=} 
		\begin{bmatrix} 
			R_\mathrm{s} + R_\mathrm{loss} & R_\mathrm{m} \\
			R_\mathrm{m} & R_\mathrm{s} + R_\mathrm{loss} \\
		\end{bmatrix} \in \mathbb{R} ^{N \times N} .
  \label{ReZ}
	\end{equation}
Within this context, the variables $R_\mathrm{s}$ and $R_\mathrm{m}$ symbolize the authentic components of self ($Z_\mathrm{m,m}$) and mutual impedances ($Z_\mathrm{m,n}$, where $m\neq n$), respectively. Subsequently, the antenna array gain can be expressed in terms of the radiation intensity and input power in the following manner:
		\begin{align}
			G(\theta,\phi) &= \frac{4\pi U(\theta, \phi)}{P_{in}} \\ 
   &= \frac{\eta}{\pi}\| {\bf{F}}(\theta,\phi)\|^2 \frac{\left|{\bf a}^H (\theta, \phi) \cdot {\bf{i}}\right|^2}{{\bf{i}}^H \cdot \Re{\{\bf{Z}\}} \cdot {\bf{i}}} \label{G1}.
		\end{align}		
Here, we assume that all dipoles in the array have an identical radiation pattern and are not influenced by variations in design features to simplify the calculation.
  
\subsection{Discussion on Realized Gain}

\begin{figure}
    \centering
    \includegraphics [scale=0.9]{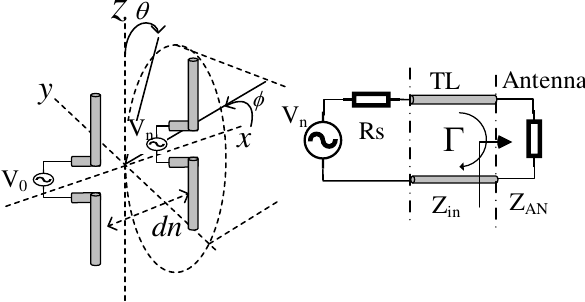}
    \caption{3-D representation of the array and equivalent circuit of the dıpole connected to a transmission line (TL).}
    \label{fig:2}
\end{figure}

In antenna engineering, transmission lines transmit the signal between the source and the antenna (see Fig. \ref{fig:2}). In an application where dipole antennas are used, coaxial transmission lines with line impedances of $50 $ $\Omega$  or $75$  $\Omega$ are generally used due to their flexible structure and easy fabrication.  In this case, necessary modifications or matching circuits should be implemented so that the impedance seen between the antenna terminal, $Z_\mathrm{in}$, is equal to $50$ $\Omega$ or $75$ $\Omega$.
Since the antenna elements are assumed not to affect each other in an uncoupled array, antenna excitations do not impact the input impedance; therefore, the self-antenna impedance, $Z_\mathrm{AN}$, is almost equal to $Z_\mathrm{in}$. In this case, adapting the antenna to the line using a balun or optimizing the antenna parameters is relatively easy. However, when mutual coupling is considered, antenna input impedances change with the feeding currents \cite{Orfanidis}. However, it is well known from the literature that some elements in coupled arrays have low input impedances. More interestingly, some parts may even have a negative input impedance \cite{SuperDirectiveAntennas}.  Hence, providing a perfect match with the line is almost impossible in such a case. As a result, a significant portion of the excitation can be reflected, reducing the efficiency of the array. 

The connection losses, commonly known as reflections (impedance mismatch) losses, are considered by incorporating a reflection efficiency (denoted as $e_\mathrm{ref}$) into the gain expression of the array. This efficiency is linked to the reflection coefficient, represented as $e_\mathrm{ref}=(1-|\Gamma|^2)$. In this context, the realized gain term ($G_\mathrm{real}=e_\mathrm{ref}G(\theta,\phi)=(1-|\Gamma|^2)G(\theta,\phi)$) can better express the effects of reflection/mismatch losses. It is evident that the realized gain decreases as the mismatch increases. Therefore, this validates that the main reason for the low realization gain in super-directive arrays is the increased impedance mismatch between the antenna element and the feed line due to mutual coupling.

In conclusion, it is crucial to articulate that achieving a high realized gain requires to minimize the mismatches. This goal can be attained by comprehensively evaluating all the structural factors influencing the antenna's radiation. Since any alteration in the antenna's structural parameters will invariably impact the reflection coefficient ($\Gamma$), it becomes imperative to carry out an optimization step for several parameters, including the antenna size, distance between the elements and the antenna feed currents.

\section{Optimization of Unit Dipole Array}
\label{sec:opt}
In this section, a DE-based optimization will be presented to design a $4$-element unit array, which is quite a challenging task in addition to two and three-element super-directive arrays in the literature \cite{ImpedanceMismatch,MultiOpt}. The number of population ($NP$) is chosen as $150$, the cross-over factor ($CR$) is $0.9$, the mutation factor ($F$) is $0.8$, and the maximum number of iterations is $250$. The following cost function is chosen to increase the realized gain
\begin{equation}
    F_\mathrm{cost}=H(\Delta|_\mathrm{RG})\Delta|_\mathrm{RG}.
    \label{eq:cost}
\end{equation}
Here, $H(.)$ and $\Delta|_\mathrm{RG}$ represent the Heaviside step function and mean squared error (MSE) between the desired ($9.16$ dB is selected, representing an approximately $40 \%$ increase compared to the uncoupled array) and calculated realized gains.


\begin{algorithm}[t]
\caption{Proposed DE-based algorithm to design a high realized four dipole element unit array.}
\begin{algorithmic}[1]
\label{alg:Alg1}
\STATE 
\textbf{Initialize}: Initialize $NP$, $CR$, $F$, number of iterations, and design parameters.\\
\STATE   {Create population}
\STATE   \textbf{Repeat: Update the population (DE/best//1/bin)}
\STATE   \hspace{0.4 em} Calculate impedance matrices.
\STATE   \hspace{0.4 em} Convert impedance matrices to scattering matrices.
\STATE   \hspace{0.4 em} Calculate combined reflection coefficients (\ref{eq:combined})
\STATE   \hspace{0.4 em} Calculate mismatch efficiency and cost function. 
\STATE   \textbf{Until:} The desired value or the maximum number of iterations has been reached.
\STATE   {Send optimized parameters to CST.}
\STATE   \textbf{Return:} Optimized currents, lengths, and positions.
\end{algorithmic}
\end{algorithm}

\begin{figure}[h]
    \centering
    \includegraphics[width=1\linewidth]{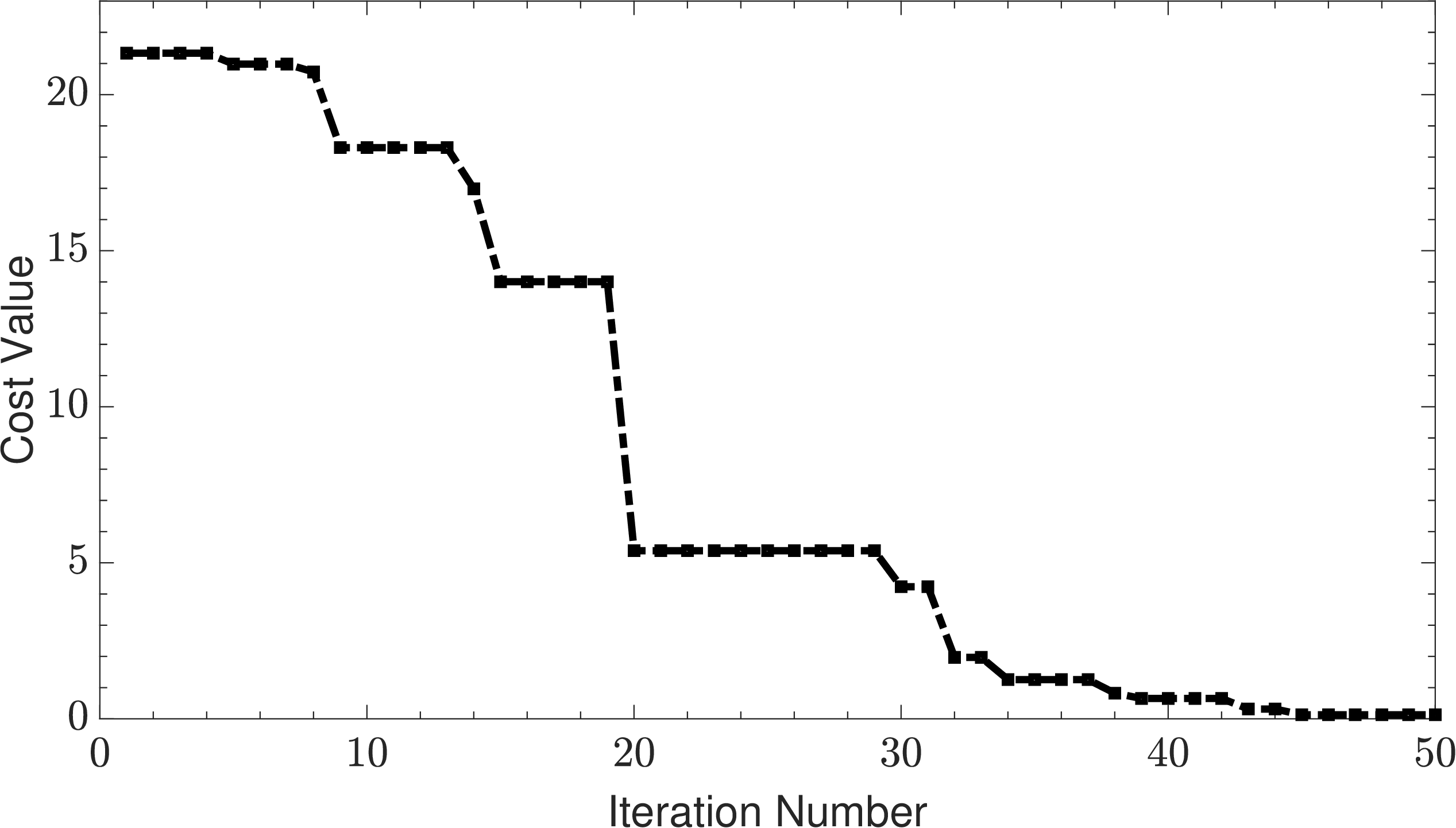}
    \caption{Convergence behaviour of the cost function in (\ref{eq:cost}). The $x$ axis is confined to 50 for best view.}
    \label{fig:cost}
\end{figure}

The optimization process is summarized in Algorithm \ref{alg:Alg1}. Initially, the algorithm begins with a predefined set of parameters ($NP, CR, F$, and number of iterations) and the design values, such as currents, lengths, and distances, are randomly selected. For each member within the population, the input impedance matrix is computed as in \cite{Orfanidis}. The transformation formulas are then applied to obtain the scattering matrix, as indicated in \cite{Pozar}. Then, using the scattering matrix, the combined active element reflection coefficient is calculated for simultaneous excitation as follows \cite{MultiCharacteristic}:

\begin{equation}
    \Gamma_\mathrm{nc}=\frac{1}{I_\mathrm{n}}\sum_\mathrm{n,m} \Gamma_\mathrm{nm}I_\mathrm{m}.
    \label{eq:combined}
\end{equation}
Here, $\Gamma_\mathrm{nc}$ represents the combined active element reflection coefficients. Subsequently, the realized gain is determined by evaluating the efficiency of the mismatch ($G_\mathrm{real}$). The algorithm assesses whether the specified target has been reached in each iteration. If the target is not achieved, the "$DE/best/1/bin$" algorithm updates the population and records the corresponding cost value. As can be seen from the cost convergence graph in Fig. \ref{fig:cost}, the algorithm started with an initial MSE of $21.33$, which is approximately equivalent to a realized gain of $5.63$ dB. Then, the targeted realized gain was successfully attained by meticulously preserving optimal the values throughout $128$ iterations.


Antenna excitation amplitudes, phases, lengths and positions obtained from the optimization are given in Table \ref{tab:my_label}. As can be seen from the table, the maximum difference between antenna amplitudes is about $0.25$ while the phase difference is about $156.3 ^\circ$.

Regarding the antenna positions, the minimum spacing between consecutive antennas becomes $9.89$ mm ($0.32 \lambda$), while the maximum spacing is $12.02$ mm ($0.40 \lambda$), both remaining below half-wavelength intervals. Furthermore, the total array size is approximately $32.26$ mm, representing a space-saving of approximately $40 \%$ compared to an uncoupled array arranged at a traditional half-wavelength inter-element distance.
Additionally, the results show that the lengths of all the elements are less than half a wavelength. 
\begin{table}[h]
    \centering
       \caption{Optimized parameters of the unit array achieved from the optimization}
        \begin{tabular}{l|c|c|c|c|}
        \hline
        \hline
       No. &  Position (mm) &  Amplitude &  Phase ($^\circ$) &  Length ($\lambda$) \\
         \hline
       1 &-16.13  & .95 & 52.47  & .44  \\
                      \hline
               2 &-6.24  & 1  & -156.37  & .45  \\
              \hline
         3 &5.78  & .96  & 0  & .45 \\
              \hline
        4 & 16.13  &  .75  & 149.11 & .48 \\
        \hline
    \end{tabular}
    \label{tab:my_label}
\end{table}

\begin{figure}[h]
    \centering
    \includegraphics[width=1\linewidth]{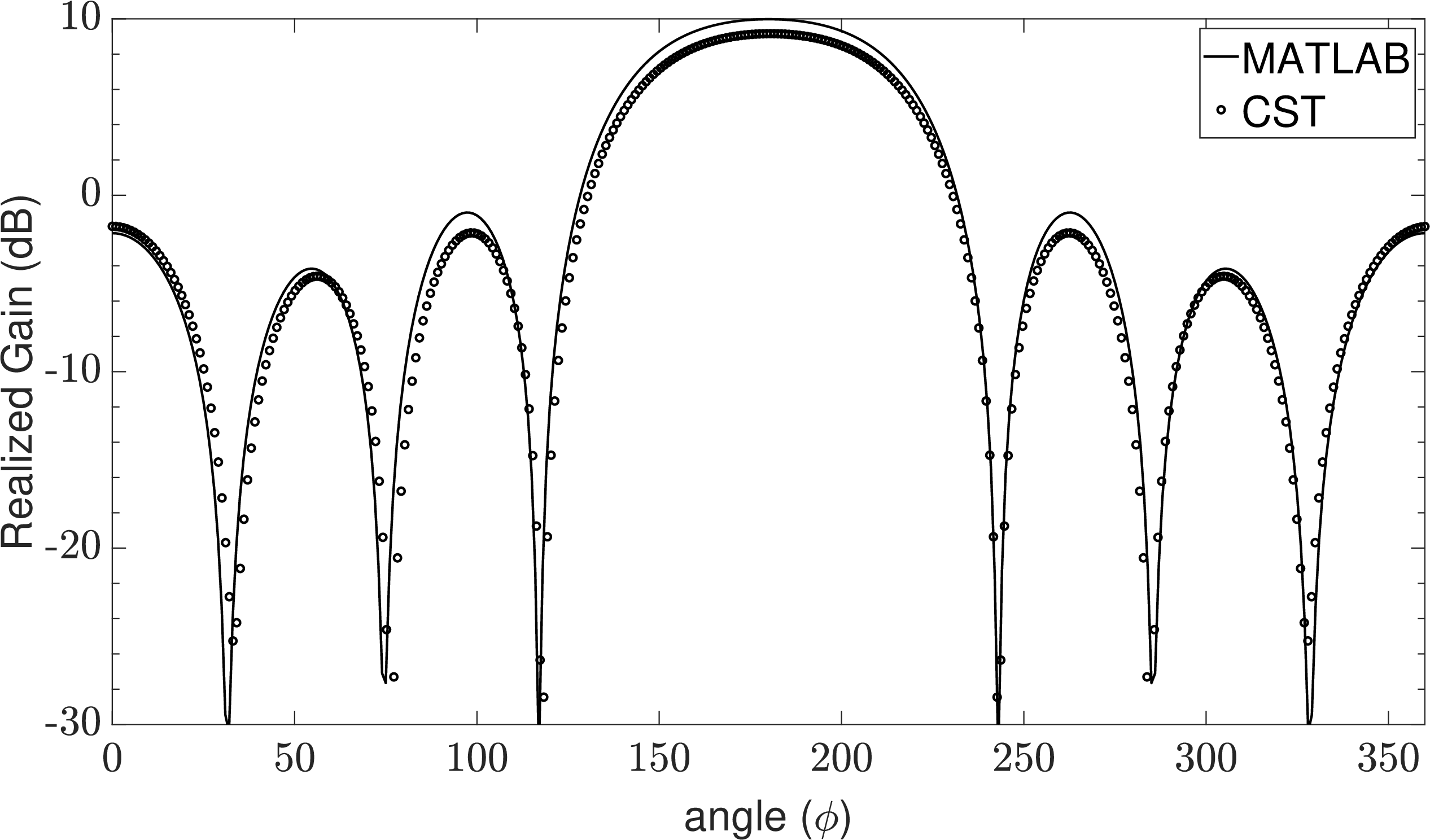}
    \caption{MATLAB and CST comparison of realized gain results obtained in the optimization.}
    \label{fig:Realizedgainfastest}
\end{figure}
The realized gain of the optimized array units is first simulated in MATLAB and then validated using CST to ensure the results' accuracy. The slight difference shown in Fig. \ref{fig:Realizedgainfastest} is because, unlike MATLAB, CST mainly uses a more realistic Method of Moments (MoM) or Finite-Difference Time-Domain (FDTD), which can calculate losses, such as surface leakage currents and edge and skin effects.
\begin{table}[h]
    \centering
       \caption{Comparison Table for Different Configurations}
        \begin{tabular}{l|c|c|c|c|c|}
        \hline
        \hline
       Conf. & $\#1$ &  $\#2$ & ULA &  Th. Exc.\cite{OptCurrent}& Optimized \\
         \hline
       R. G. (dB) &7.89 & 6.37 & $8.95 \footnotemark$   &7.60 & 9.16 \\
                      \hline
              Tot. Eff. ($\%$) & 57.83 & 84.42& 94.95 & 54.00 & 80.11\\
              
        \hline
    \end{tabular}
    \label{tab:comparison}
\end{table}

\footnotetext{Recall that in classical ULA, the absence of phase differences between elements results in maximum radiation in the broadside direction, while super-directive arrays exhibit maximum radiation towards the end-fire.}

After confirming the results through numerical validation, we explored three distinct array arrangements as detailed in Table \ref{tab:comparison} to underscore the advantages of the optimized configuration. In the initial scenario, denoted as $\#1$, the lengths of the elements are maintained at a constant half-wavelength, unlike the optimized array. Significantly, the realized gain outperforms that of configuration $\#1$ by approximately $32\%$. Transitioning to configuration $\#2$, where a uniform half-wavelength inter-element spacing was employed, the optimized array yields an increase in the realized gain of approximately $90\%$.
In our comparison, the third configuration involves a uniformly linear array (ULA) with half-wavelength elements and inter-element distances. In this case, the optimized array demonstrated a realized gain approximately $4\%$ higher. It is crucial to emphasize that in a ULA, the maximum radiation is directed symmetrically to the $x$-plane in the broadside direction ($\phi=90^\circ$). However, it is widely recognized that super-directive arrays achieve peak radiation in the end-fire direction \cite{CompEnd,Maximumdirectivity}. When we introduce the necessary phase adjustments to orient the ULA towards the end fire, the resulting gain is $5.82$ dB. These results signify that the optimized array surpasses the ULA's realized gain by more than $3.34$ dB.
Furthermore, maintaining a consistent overall array size along the $x$-axis as the optimized array, employing a uniform distance of approximately $0.35\lambda$ between elements, and utilizing current values from existing literature for array excitation result in the optimized array achieving a realized gain approximately $43\%$ higher.

\section{Analysis of Linear Sub-array Configurations}
\label{sec:subarray}
In this section, the linear array configurations of the optimized array are analyzed. Each unit of the array is arranged along the $x$-axis with the predetermined distance between them. Therefore, the magnitude of the total electric field at $\theta=90^\circ$ for this array, consisting of $S$ sub-arrays, is expressed as follows:
\begin{equation}
    E(r,\theta, \phi)=-j\eta \frac{e^{-jkr}}{2\pi r} \sum_{s=0}^{S-1} \sum_{n=0}^{N-1}I_\mathrm{n} e^{jk(d_\mathrm{n}+s(D_\mathrm{g}+d_\mathrm{N-1}))\cos\phi}.
    \label{eq:subarray}
\end{equation}
Here, $D_g$ represents the distance between the unit arrays, and $d_n$ denotes the positions of the elements on the $x$-axis. Thus, (\ref{eq:subarray}) can be rewritten as follows:

\begin{equation}
    E(r,\phi)=-j\eta \frac{e^{-jkr}}{2\pi r} \sum_{s=0}^{S-1} e^{jks((D_\mathrm{g}+d_\mathrm{N-1})cos\phi)} \sum_{n=0}^{N-1} e^{jkd_\mathrm{n}cos\phi},
\end{equation}
where the first summation represents the group response, and the second summation represents the unit array element response. If $D_\mathrm{g}$ is selected to minimize the inter-group mutual couplings (more than half wavelength in theory but more than one wavelength in practice), the total input power can be written by following (\ref{eq:Pin}) as:
\begin{equation}
   P_\mathrm{in}^\mathrm{array}= S\frac{1}{2} {\bf{i}}^H \cdot \Re{\{\bf{Z}\}} \cdot {\bf{i}} .
\end{equation}
Here, ${\bf Z}  \in {\mathbb C} ^{N \times N}$ is the unit dipole array impedance matrix. Therefore, the gain of the linear array can be calculated by following (\ref{ReZ}) and (\ref{G1}).
\subsection{Numerical examples for linear array }
This section shows numerical examples of the unit array converted to a linear two-subarray configuration with a 2-wavelength group distance. In this case, $NS=8$ since the total number of elements is $N=4$, $S=2$.
\begin{figure}[h]
    \centering
    \includegraphics[width=1\linewidth]{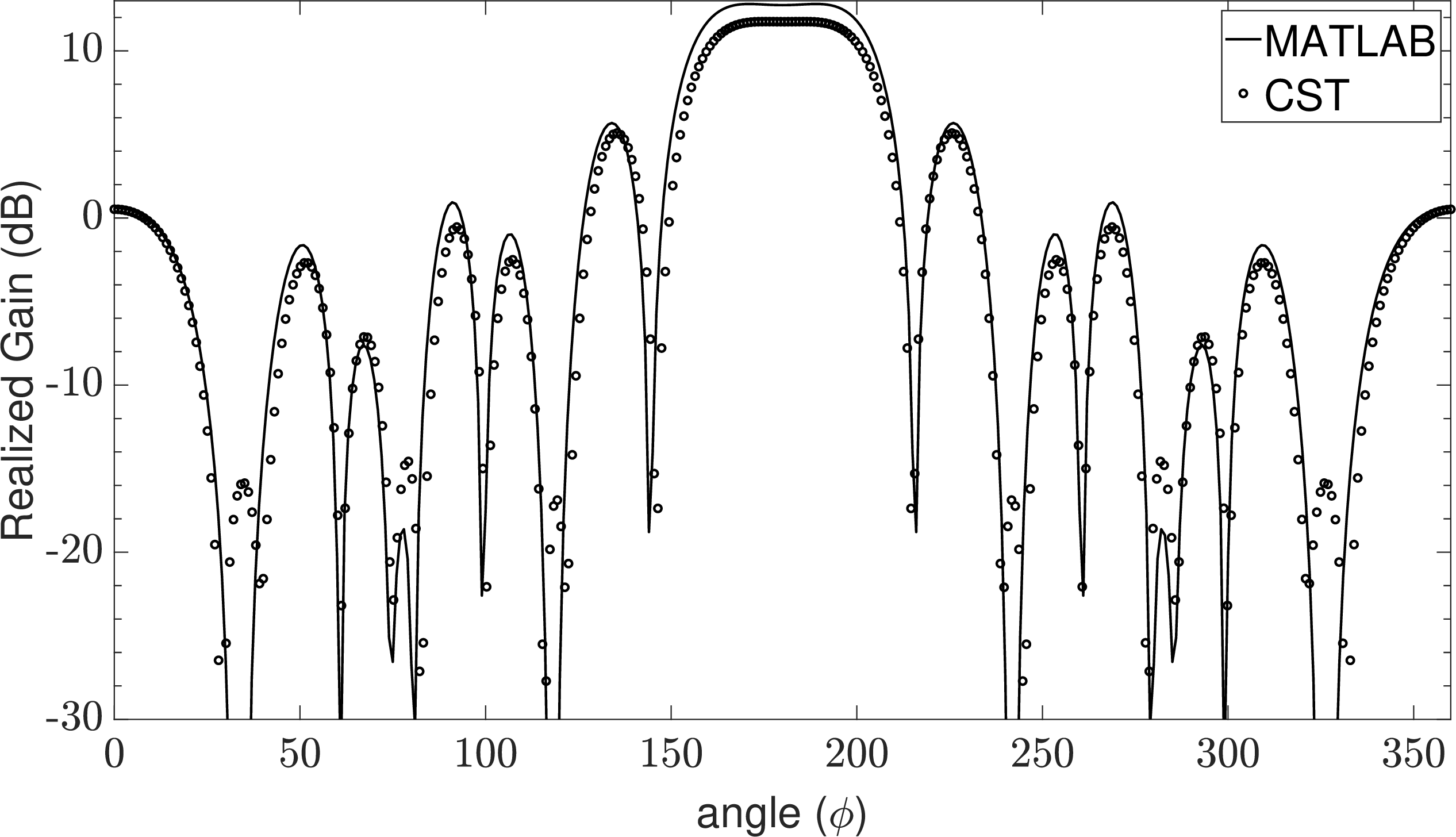}
    \caption{MATLAB and CST comparison of realized gain results for linear configuration obtained in the optimization.}
    \label{fig:Realizedgainfastestarray}
\end{figure}

The realized gain-$\phi$ graph obtained is given in Fig. \ref{fig:Realizedgainfastestarray}. As can be seen from the figure, the MATLAB and CST results are very close to each other. Therefore, it can be inferred that MATLAB can serve as an alternative to high-computing hardware methods like MoM or FDTD for optimization.

\section{Conclusion}
\label{sec:conclusion}
In this study, we investigated the mismatch between the feed line and antenna elements, a critical factor contributing to the low realized gain in super-directive arrays. Our study highlighted that relying on a single matching circuit proves inadequate when the antenna elements are strategically arranged to couple. As a result, ensuring the compatibility of all elements with the transmission line becomes a complex optimization task, necessitating the consideration of various design parameters, such as feeding currents, lengths, and positions.

We utilized the DE algorithm to optimize a 4-element dipole array to address this issue, resulting in a significantly improved realized gain. Once the targeted realized gain value was achieved, we thoroughly examined and verified the sub-array configurations of the unit dipole array using a full-wave electromagnetic simulation program.

The results demonstrated that the proposed unit array has excellent characteristics, with high gain and efficiency parameters, making it a suitable candidate for dense array applications, particularly in massive MIMO topologies \cite{6G}. These findings highlight the potential for substantial improvements in super-directive arrays by addressing the mismatch issues and optimizing the array design using the DE algorithm.
\bibliographystyle{IEEEtran} 				
\bibliography{referans.bib}

\end{document}